\documentclass[aip,cha,reprint]{revtex4-1}
\usepackage{amsmath}
\usepackage{amssymb}
\usepackage{units}
\usepackage{graphicx}
\usepackage{xcolor}
\usepackage{soul}

\newcommand{\w}{\omega}

\newcommand{\e}{\varepsilon}

\begin{document}

\title{Controlling collective synchrony in oscillatory ensembles by precisely timed pulses
} 

\author{Michael Rosenblum}
\email{mros@uni-potsdam.de}
\affiliation{Institute of Physics and Astronomy, University of Potsdam,
Karl-Liebknecht-Str. 24/25, 14476 Potsdam-Golm, Germany}

\date{\today}

\begin{abstract}
We present an efficient technique for control of synchrony in a globally coupled ensemble by 
pulsatile action. We assume that we can observe the collective oscillation and can stimulate 
all elements of the ensemble simultaneously.
We pay special attention to the minimization of intervention into the system. 
The key idea is to stimulate only at the most sensitive phase. To find this phase we implement 
an adaptive feedback control. Estimating the instantaneous phase of the collective mode on 
the fly, we achieve efficient suppression using a few pulses per oscillatory cycle.
We discuss the possible relevance of the results for neuroscience, namely for the development 
of advanced algorithms for deep brain stimulation, a medical technique used to treat Parkinson's disease.
\end{abstract}

\maketitle 

\begin{quotation}
Networks of highly-interconnected oscillatory elements are popular models for various systems, 
either manufactured or natural.  It is well-known that, for sufficiently strong interaction,  
the units of the network synchronize, and the system as a whole exhibits a collective rhythm. 
Frequently this rhythm is detrimental and shall be suppressed: the examples include oscillation 
of pedestrian bridges and some pathological brain activity. On the contrary, if the interaction 
within the network is too weak to induce collective oscillation, enhancement of synchrony may 
be desirable, e.g., to ensure coherent oscillation of many low-power sources so that they produce 
a high-power output.
These two related problems call for efficient control techniques, and various schemes have been 
designed for this purpose. Here we elaborate on a special case when the control action shall be 
pulsatile, which is a common requirement for neuroscience applications. 
We develop a feedback-based adaptive technique that achieves suppression of undesired collective 
synchrony with only one or two pulses per oscillation cycle. 
A slightly modified version of this technique enhances 
collective synchrony if required. We discuss a possible application to a clinical technique, 
deep brain stimulation, widely used to treat several neurological diseases.
\end{quotation}

\section{Introduction}
\label{intro}
The nonlinear science community has paid a lot of attention to research on large populations of interacting 
self-oscillatory units.  Hundreds (if not thousands) of research articles followed the pioneering publications 
on this topic~\cite{Winfree-67,*Winfree-80,Kuramoto-75,*Kuramoto-84}. 
Many of them exploited the analytically tractable model of globally coupled phase
oscillators~\cite{Kuramoto-75,*Kuramoto-84}. 
Theoretical, numerical, and experimental studies described and analyzed many interesting phenomena. 
An incomplete list includes the emergence of the collective mode, clustering, quasiperiodic dynamics, 
appearance of heteroclinic cycles, and chimera states, see 
reviews~\cite{Strogatz-00,*Strogatz-03,Pikovsky-Rosenblum-Kurths-01,*Pikovsky-Rosenblum-15,%
Acebron-etal-05,*Osipov-Kurths-Zhou-07,*Breakspear-Heitmann-Daffertshofer-10}
and references therein. 

The most important and most studied effect is the emergence of the collective oscillation in the population 
due to the synchronization of individual units. Collective synchrony can be important for maintaining 
high-power output in a population of low-power generators and is known to play a significant role in the 
generation of both vital and pathological biological rhythms. 
Therefore, control of synchrony, i.e. either suppression or enhancement of the collective 
mode, is a challenging problem. 
In particular, the suppression task is motivated by a possible relevance to a widely used clinical procedure, 
deep brain stimulation (DBS). DBS implies high-frequency pulse stimulation of some brain areas and aims at an 
improvement of motor symptoms in Parkinsonian patients as well as in the case of some other 
pathologies~\cite{Benabid_et_al-91,*Benabid_et_al-09,*Kuehn-Volkmann-17}. 
Though the mechanisms of DBS remain in the focus of research in 
neuroscience~\cite{Johnson2008,*Gradinaru-09,*Deniau_et_al-10}, many researchers from the 
nonlinear science community have adopted a working hypothesis that views DBS 
as a desynchronization task~\cite{Tass-99,*Tass-00,*Tass-01,*Tass_2001,*Tass-02}. 
This hypothesis has been exploited in a number of model studies suggesting open-loop 
and closed-loop techniques for 
suppression~\cite{Rosenblum-Pikovsky-04,*Rosenblum-Pikovsky-04a,%
Popovych-Hauptmann-Tass-05,%
Tukhlina-Rosenblum-Pikovsky-Kurths-07,Hauptmann-Tass-09,*Popovych-Tass-12,Montaseri_et_al-13,%
Lin_2013,*Zhou_2017,*Wilson-Moehlis-16,*Holt_et_al-16,Popovych_et_al-17,Krylov-Dylov-Rosenblum-20}.  
In this paper, we follow this line of research and consider both the suppression and the enhancement 
task for a globally coupled network. 
We extend our previous studies on feedback-based 
control~\cite{Rosenblum-Pikovsky-04,*Rosenblum-Pikovsky-04a,
Tukhlina-Rosenblum-Pikovsky-Kurths-07,Montaseri_et_al-13,Popovych_et_al-17},
concentrating on the case of pulsatile 
stimulation. With the goal to minimize the intervention into the controlled system, we employ precisely 
timed pulses, applied at a vulnerable phase that is determined on the fly. In this way, we efficiently 
desynchronize the oscillatory activity by a few pulses per oscillatory cycle.

The paper is organized as follows. In Section~\ref{MainIdea} we present the simplest model of globally coupled 
Bonhoeffer -- van der Pol oscillators and use it to introduce and illustrate the main idea. 
Here we also discuss how the phase of the collective oscillation can be obtained in real-time. 
In Section~\ref{sec:auto} we present the algorithm for adaptive tuning of the feedback parameters 
and illustrate 
its performance with the help of the ensemble of chaotic R\"ossler oscillators. 
Section~\ref{sec:chbal} takes into account limitations inherent to neuroscience and presents suppression by the 
so-called charge-balanced pulses. Section~\ref{sec:enh} is devoted to the enhancement of collective 
synchrony while Section~\ref{sec:sum} summarizes and discusses the results.


\section{Pulses applied at a vulnerable phase}
\label{MainIdea}
\subsection{The basic model and the main idea}
We introduce the approach using as an example a simple model of $N$ globally 
coupled  Bonhoeffer--van der Pol oscillators:
\begin{equation}
\begin{cases}
\dot{x}_k &= x_k-x_k^3/3 - y_k +I_k +\e X + \cos\psi\cdot P(t)\;,\\
\dot{y}_k &= 0.1 (x_k-0.8y_k+0.7)+\sin\psi\cdot P(t)\;,
\end{cases}
\label{eq:bvdp}
\end{equation}
where $k$ is the oscillator index, $k=1,\ldots,N$, and the term $\e X$ describes the global coupling. Here
$X$ is the mean field, $X=N^{-1}\sum_k x_k$, and the coupling coefficient $\e$ explicitly describes 
the interaction between the elements of the ensemble. 
The oscillators are not identical: their frequencies are determined by the parameter $I_k$ 
that is Gaussian-distributed with the mean $0.6$ and standard deviation $0.1$. 
$P(t)$ is external pulsatile action 
applied to the ensemble; it will be specified below. 
Finally, the parameter $\psi$ describes how the external pulses act on the system.
This parameter is considered to be unknown, to imitate the uncertainty in stimulation 
of a real-world system without any knowledge of its model. 
 
\begin{figure}
\centering
\includegraphics[width=0.75\columnwidth]{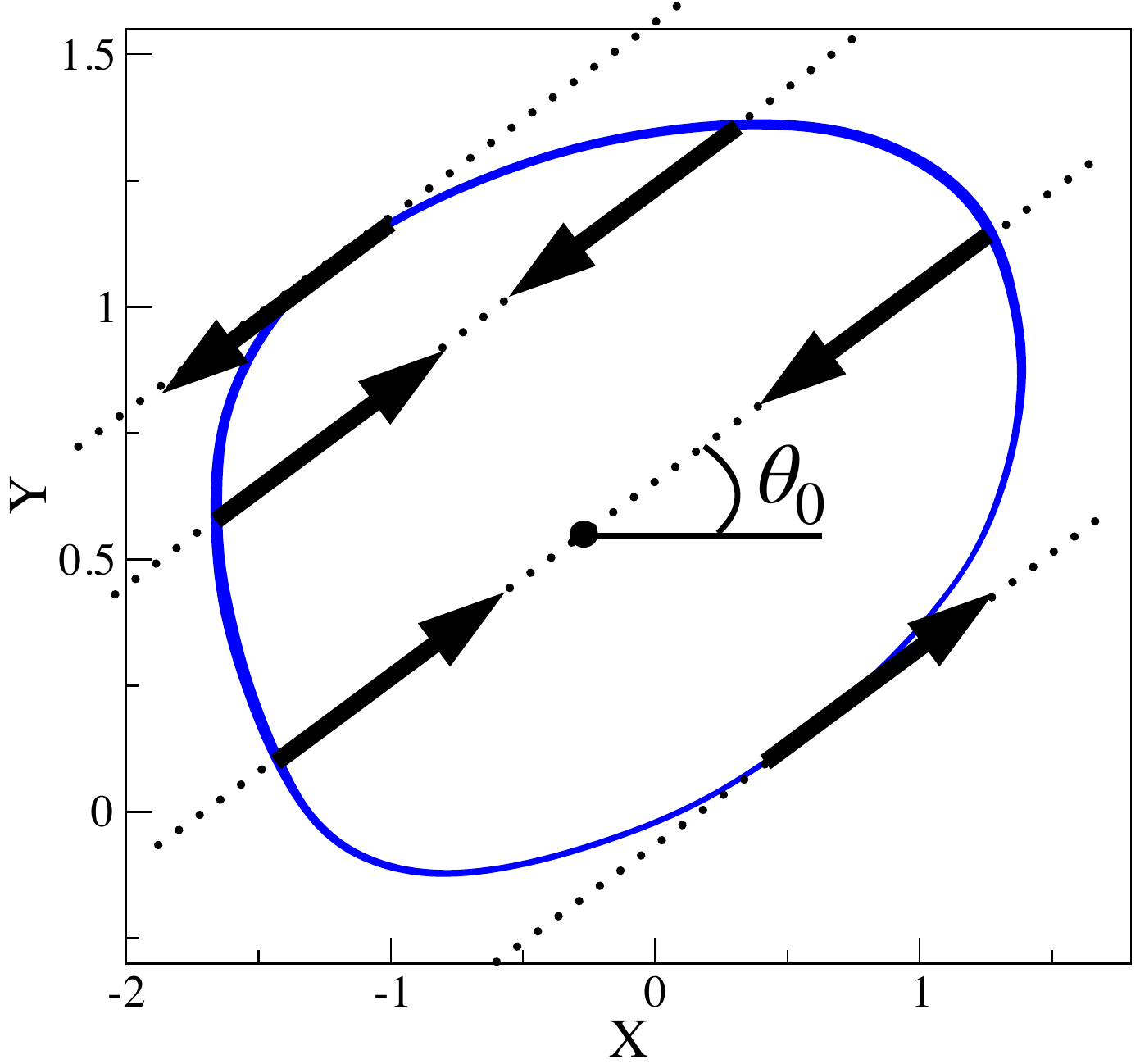}
\caption{Qualitative explanation of the approach. Blue solid line is the 
limit cycle of the collective mode of the system~(\ref{eq:bvdp}) 
(the collective oscillation is periodic, up to finite-size fluctuations).  
Suppression of the collective mode can be achieved if the applied pulse pushes
the system towards the unstable fixed point, shown by a small filled
circle at $X\approx -0.27\,,Y\approx 0.55$. 
The direction of the applied pulses cannot be chosen: it is predetermined 
by the equations of the systems and by the way the stimulation enters these
equations; this direction is shown by dotted lines. Obviously, 
the oscillation amplitude is mostly affected by a pulse, applied when the 
system's state is close to the phase angle $\theta_0$ or $\theta_0+\pi$, and less
affected if the phase angle is close to $\theta_0\pm \pi/2$. 
Thus, efficient suppression can be achieved by a repetitive application 
of pulses of certain polarity at about $\theta_0$ and of inverse polarity
at $\theta_0+\pi$.
}
\label{fig1}
\end{figure}

Figure~\ref{fig1} illustrates the dynamics of the autonomous ensemble, $P(t)=0$; 
here we plot $Y=N^{-1}\sum_k y_k$ vs. $X$ for $\e=0.03$ and $N=1000$. 
A symbol at $X\approx -0.27\,, Y\approx 0.55$ shows the unstable fixed point of the globally coupled 
system~\footnote{The fixed point can be found by simulating the ensemble for $\e=0$.}.
In this representation, suppression of the collective oscillation $X(t)$ means that
the system is put into and is kept in a vicinity of the unstable fixed point.  
Suppose the applied external pulses act along a certain direction, indicated by dashed 
lines in Figure~\ref{fig1}.
Obviously, the pulses applied to the system at phase angles close to $\theta_0$ and the pulses 
of an opposite polarity applied at approximately $\theta_0+\pi$ are most efficient 
for reducing the collective oscillation, and, hence, for desynchronization. 
On the contrary, the oscillation amplitude is much less affected by 
the pulses applied around $\theta_0\pm \pi/2$.  
This qualitative discussion presents the main idea of our approach: in order to achieve the
control goal with minimal intervention we have to stimulate only in a small interval around the vulnerable 
phase $\theta_0$.~\footnote{We emphasize that phase angle $\theta$ is not the true 
phase of the self-sustained oscillatory system but only 
a protophase, see \cite{Kralemann_et_al-07,*Kralemann_et_al-08}, 
but this distinction is not important for our problem.
We also stress, that $\theta$ is related to the parameter $\psi$
but is not equal to it, as will be discussed below.
}
For the rest of this Section we assume that $\theta_0$ is known, while in Section~\ref{sec:auto} 
we drop this assumption and show how $\theta_0$ can be found.

\subsection{Phase estimation}
For efficient stimulation, we have to monitor the 
instantaneous phase of the collective oscillation on the fly,
assuming that we observe only a scalar time series. 
Below we suppose that $X(t)$ is measured. 
To this end, we follow
\cite{Tukhlina-Rosenblum-Pikovsky-Kurths-07} and introduce a ``device'' consisting of 
a harmonic linear oscillator and an integrating unit
\begin{align}
&\ddot{u}+\alpha\dot u +\w_0^2 u = X(t)\;,\label{eq:filter}\\
&\mu\dot{d} + d = \dot u \;.              \label{eq:integrator}
\end{align}
The role of the harmonic oscillator Eq.~(\ref{eq:filter}) is twofold. 
First, it acts as a band-pass filter
and extracts the oscillatory mode of our interest from its mixture with noise. 
Second,
it yields signal $\dot u$ which phase is close to that of the input $X(t)$, 
provided the frequency $\w_0$ is chosen to be close to the mean frequency of $X(t)$.
The integrating unit Eq.~(\ref{eq:integrator}) provides a signal, shifted by $\pi/2$ 
with respect to $\dot u$. 
It is convenient to introduce two auxiliary variables $\hat x=\alpha\dot u$ and 
$\hat y=\alpha\w_0\mu d$; their amplitudes are close to that of $X(t)$ while 
their phases are delayed by $0$ and $\pi/2$, respectively, 
cf.~\cite{Tukhlina-Rosenblum-Pikovsky-Kurths-07,Montaseri_et_al-13}. Hence, 
we can estimate the instantaneous (proto)phase of $X(t)$ as 
\begin{equation}
\theta(t)=\arctan(\hat y/\hat x)\;.
\label{eq:phase}
\end{equation} 
In the following, we will also need the instantaneous amplitude
\begin{equation}
a(t)=\sqrt{\hat x^2 +\hat y^2}\;.
\label{eq:amplitude}
\end{equation} 

\begin{figure}
\centering
\includegraphics[width=0.9\columnwidth]{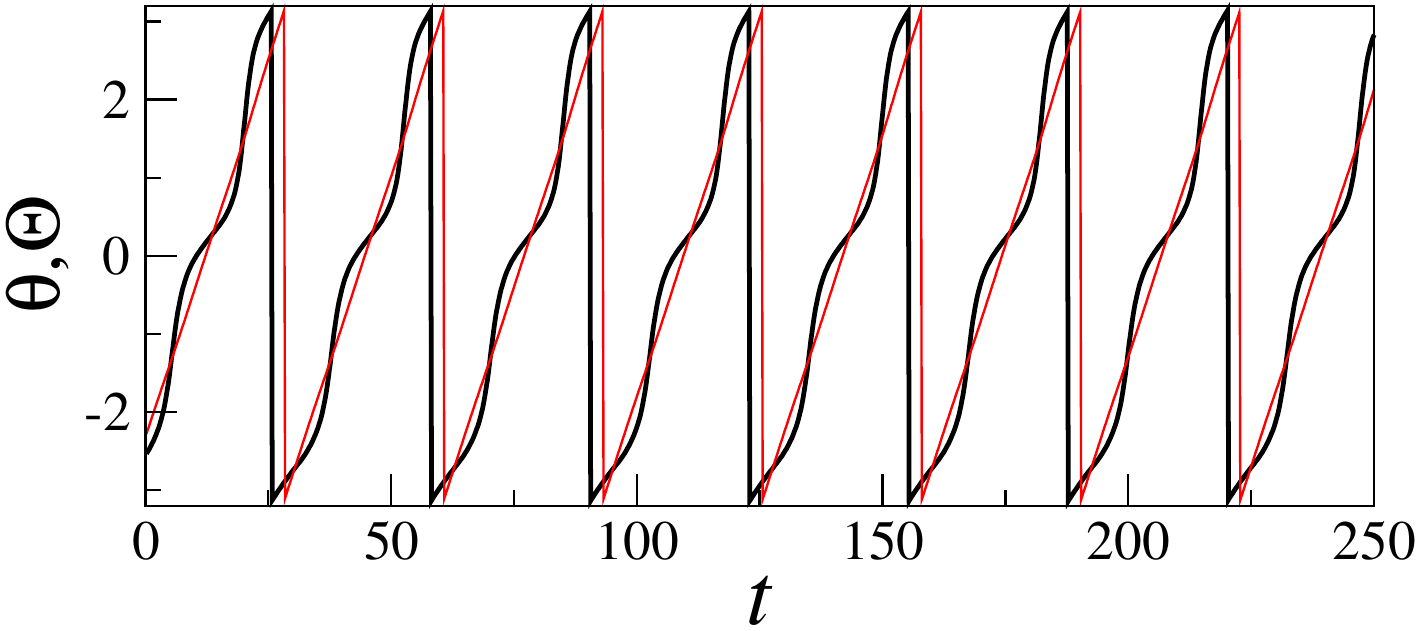}
\caption{Estimation of the phase of the collective mode of the autonomous 
ensemble Eq.~(\ref{eq:bvdp}) 
with the help of the ``measuring device'' described by Eqs.~(\ref{eq:filter},\ref{eq:integrator}).
Red line shows phase $\theta$ computed according to Eq.~(\ref{eq:phase}) while 
the black bold line shows the angle variable $\Theta=\arctan[(Y-0.55)/(X+0.27)]$.
We emphasize that this algorithm does not require to know the future values of the signal, as
is required if, e.g., the Hilbert transform is used. In other words, in this way $\theta$ 
is obtained in real time. 
}
\label{fig2}
\end{figure}
Figure~\ref{fig2} illustrates how the algorithm for phase estimation works with the 
system~(\ref{eq:bvdp}). The parameter values used here are:
$\w_0=2\pi/32.5$, $\alpha=0.3\w_0$, and $\mu=500$.

\subsection{Timing and strength of stimuli}
To determine when and how to stimulate, we trace the instantaneous phase 
$\theta(t)$ and check whether
\begin{equation}
|\theta(t)-\theta_0|<\Theta_{tol}\quad\text{or}\quad |\theta(t)-\theta_0-\pi|<\Theta_{tol}\;.
\end{equation}
If one of these conditions is fulfilled at time instant $t_n$ then a pulse of a certain 
strength $A_n$ is applied to all elements of the ensemble.
Here $\Theta_{tol}$ is the tolerance parameter. For a fixed width of stimulation pulses 
$\Theta_{tol}$ determines whether one pulse (if $\Theta_{tol}$ is small) or several 
pulses (if $\Theta_{tol}$ is sufficiently large) are applied around $\theta_0$
or $\theta_0+\pi$, respectively.
The strength of each pulse, $A_n$, is limited by the maximal allowed value, 
$|A_n|\le A_0$, and is determined by the current value of the 
instantaneous amplitude $a(t_n)$: 
\begin{equation}
A=\pm\text{max}(\e_{fb}a(t_n),-A_0)\;,
\label{eq:pstrength}
\end{equation}
where positive and negative signs correspond to stimulation around $\theta_0$ and 
$\theta_0+\pi$, respectively.
Here $\e_{fb}<0$ is the strength of the negative feedback. 

\subsection{A numerical example}

\begin{figure}
\centering
\includegraphics[width=0.65\columnwidth]{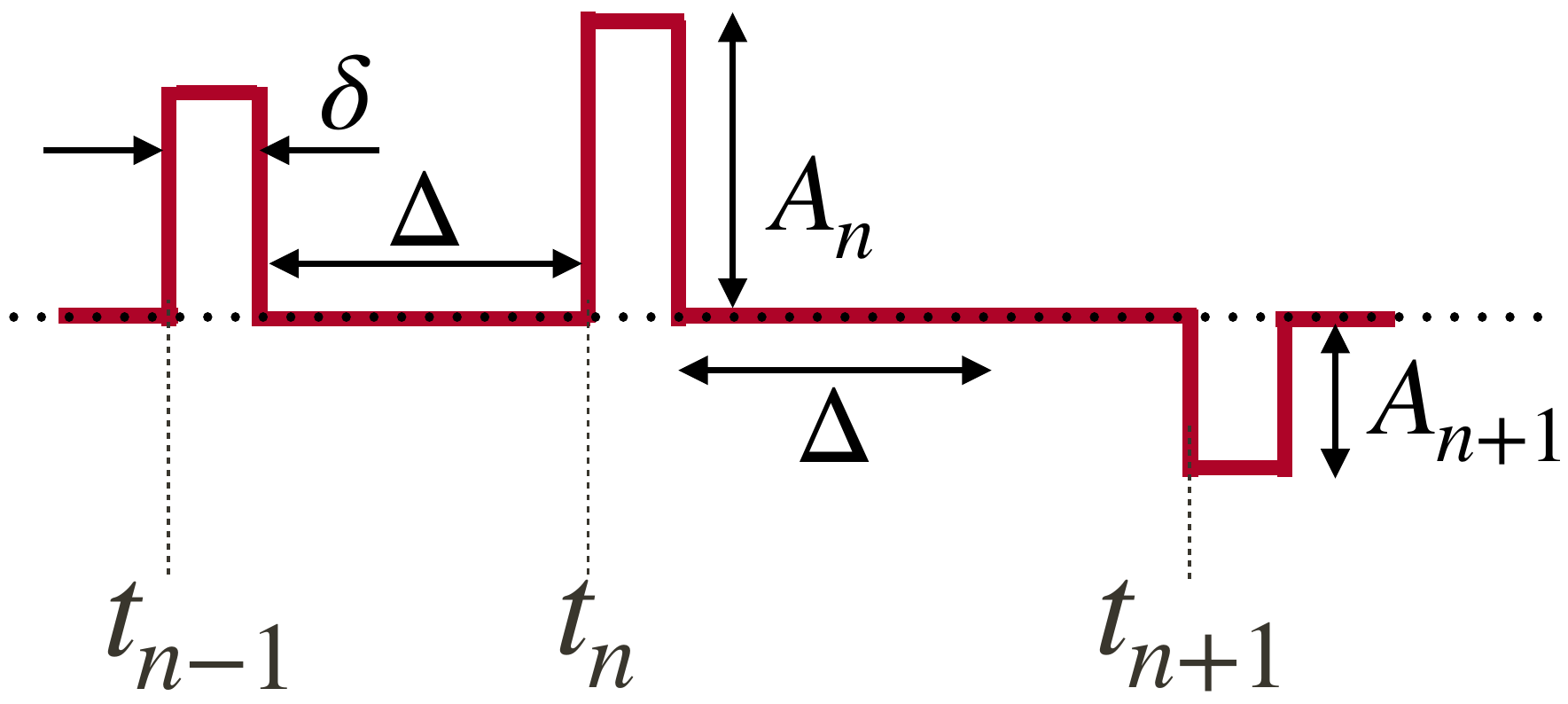}
\caption{Stimulation by rectangular pulses. 
Fixed parameters $\delta$ and $\Delta$ determine the pulse width and the minimal 
inter-pulse interval, respectively.
The pulse amplitude, $A_n$, varies from pulse to pulse, as determined 
by Eq.~(\ref{eq:pstrength}) according to the instantaneous amplitude $a$ 
at time instant $t_n$. Notice that  $A_n$ can be both positive and negative.
}
\label{fig:pulse_shape}
\end{figure}

\begin{figure}
\centering
\includegraphics[width=0.75\columnwidth]{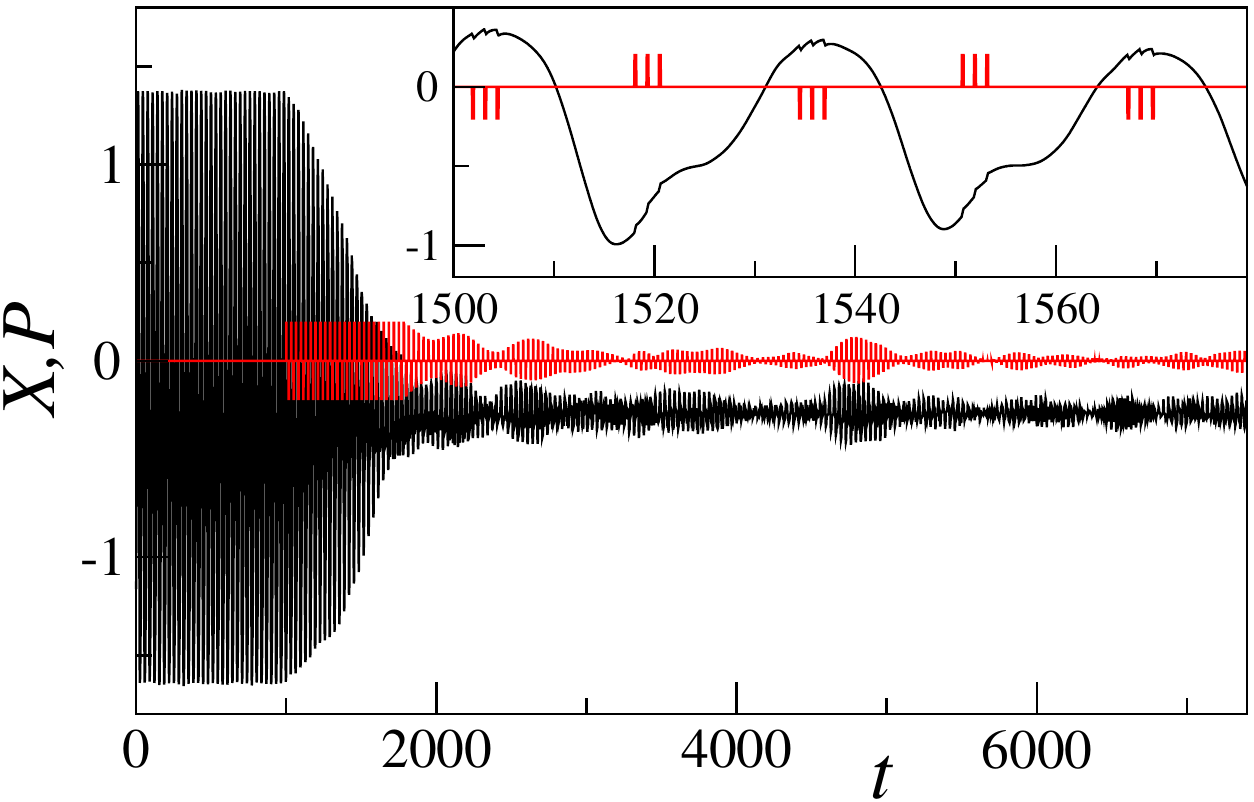}
\caption{Suppression of the collective mode in system~(\ref{eq:bvdp}) 
by rectangular pulses. The stimulation is switched on at $t_0=1000$. 
Pulse width and minimal inter-pulse distance are $\delta=0.2$ and $\Delta=1$,
respectively (to be compared with the period of the collective 
oscillation $T\approx 32.5$).
}
\label{fig:psi0}
\end{figure}

For the first illustration of the approach, 
we consider the model (\ref{eq:bvdp}) with $N=1000$ and $\e=0.03$
and try to suppress the collective oscillation by rectangular pulses 
of the constant width $\delta$ and minimal inter-pulse interval $\Delta$, 
see Fig.~\ref{fig:pulse_shape}.
We set $\psi=0$,  and stimulate with negative pulses 
around $\theta_0=0$ and with positive pulses around $\pi$.
Other parameters are $\e_{fb}=-0.05$, $\Theta_{tol}=0.08\pi$, and $A_0=0.2$.
Figure~\ref{fig:psi0} demonstrates efficient suppression of the
collective oscillation. For the chosen $\Theta_{tol}$ there are three
(sometimes four) stimuli in a bunch around $\theta_0$ or $\theta_0+\pi$.

\begin{figure}
\centering
\includegraphics[width=0.9\columnwidth]{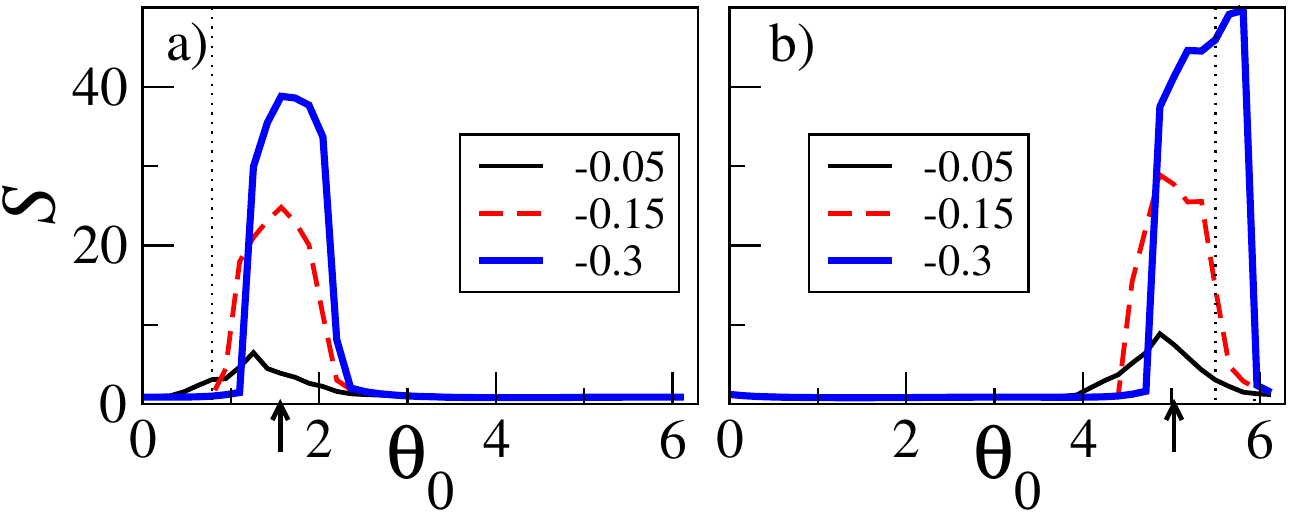}
\caption{Suppression coefficient for system~(\ref{eq:bvdp}) as a function 
of $\theta_0$ for $\psi=\pi/4$ (a) and $\psi=-\pi/4$ (b) and three values 
of the feedback strength, $\e_{fb}=-0.05, -0.15, -0.3$.
$\Theta_{tol}=0.08\pi$, $A_0=0.2$, 
$\delta=0.2$, and $\Delta=1$. Vertical arrows touching the horizontal axis indicate
the optimal values of $\theta_0$ detected by an automated algorithm described 
in Section~\ref{sec:auto}.
}
\label{fig:psi_theta}
\end{figure}

Before proceeding with the further details of our approach, we discuss the meaning of the 
\textit{a priori} unknown parameter $\psi$. It describes the distribution of the stimulation 
between the equations and is related to phase shift, inherent to stimulation. The latter
also depends on the property of individual oscillators and of the coupling between them,
see a discussion in \cite{Tukhlina-Rosenblum-Pikovsky-Kurths-07} and references therein.
Thus, $\psi$ is related to $\theta_0$, though is not exactly equal to it.
To illustrate this and to analyse sensitivity of our technique to the choice of $\theta_0$
we compute the suppression coefficient $S$ as a function of $\theta_0$, for $\psi=\pm\pi/4$ 
(Fig.~\ref{fig:psi_theta}). 
The coefficient is determined as 
\begin{equation*}
S=\text{std}(X)/\text{std}(X_s)\;,
\end{equation*}
where $\text{std}$ means standard deviation and 
$X$ and $X_s$ are the mean fields in the unstimulated and stimulated system, respectively. 
The results show that choice of $\theta_0$ is crucial and therefore we need a technique 
for tuning $\theta_0$ as well as the feedback strength $\e_{fb}$ automatically. 
This technique is presented in the next Section.

\section{Automatic tuning of suppression parameters}
\label{sec:auto}
For a proper tuning of the feedback-based suppression algorithm we adapt 
the approach developed in our previous publication~\cite{Montaseri_et_al-13}.
Namely, we adjust parameters $\theta_0$, $\e_{fb}$ after each complete cycle, according to the 
averaged value $\bar a$ of the instantaneous amplitude $a(t)$, see Eq.~\ref{eq:amplitude}.
To be exact, the latter is averaged over all points within one cycle, 
except for the interval
where the system is stimulated (i.e. except for the points 
where $|\theta-\theta_0|<\Theta_{tol}$ and
$|\theta-\theta_0-\pi|<\Theta_{tol}$). 
The update rules are
\begin{align}
&\theta_0\;\to\; \theta_0+k_1\bar a(1+\tanh[k_2(\bar a-a_{stop})]\;,\label{adaptthe}\\
&\e_{fb}\;\to\; \e_{fb}-k_3\bar a/\cosh(k_4\e_{fb})\;,  \label{adapteps}
\end{align}
where $k_i$ and $a_{stop}$ are parameters.
The initial conditions, if not said otherwise, are $\theta_0(t_0)=0$, $\e_{fb}(t_0)=0$. 

\begin{figure}
\centering
\includegraphics[width=0.9\columnwidth]{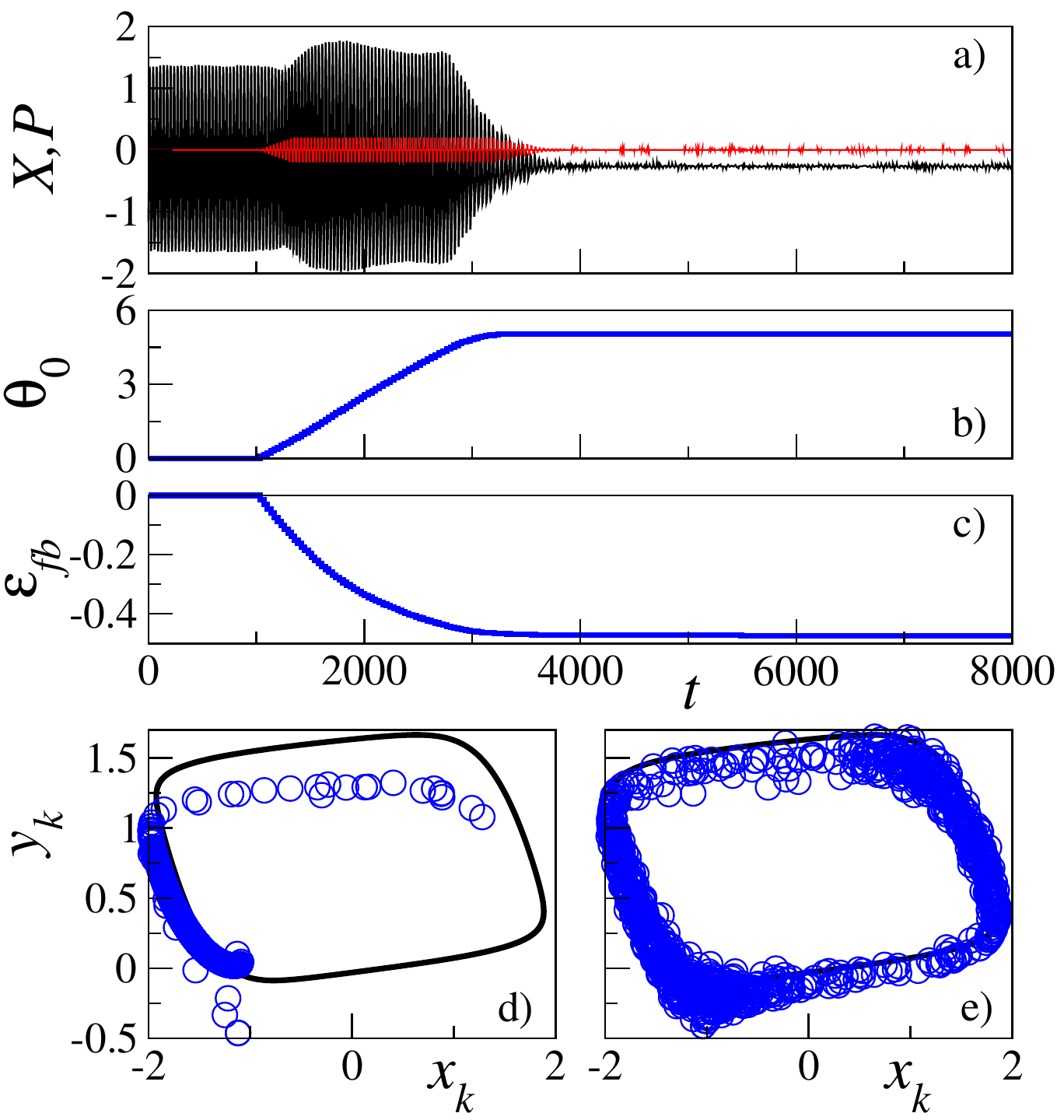}
\caption{Suppression of the collective mode in system~(\ref{eq:bvdp}) 
by an adaptive technique. Panel (a) shows the mean field and stimulation that 
is smoothly switched on at $t_0=1000$. Panels (b,c) show the time evolution of 
two feedback parameters, $\theta_0$ and $\e_{fb}$ that vary unless the mean field
$X$ is suppressed. Panels (d,e) show snapshots of the ensemble in the synchronous 
state (before the feedback is switched on) and after suppression is achieved, 
respectively. The snapshots demonstrate that in the desynchronized state 
the individual units continue to oscillate, though not coherently.
}
\label{fig:bvdpadapt}
\end{figure}

An example of suppression with an automated tuning of parameters is illustrated in 
Fig.~\ref{fig:bvdpadapt}, for $\psi=-\pi/4$. 
We see that detected value of $\theta_0$ here is $\theta_0\approx 5.03$, cf. Fig.~\ref{fig:psi_theta}a; 
the suppression factor is $S=52.6$. 
Stimulation is turned on smoothly and its onset is followed by a temporal increase of synchrony, because $\theta_0$
is swept through the interval of angles that are beneficial for enhancement.
For $\psi=\pi/4$ (not shown) the transient is shorter and there 
is no intermediate increase in the amplitude of the mean field. 
In the desynchronized state $S=37.7$ and 
$\theta_0\approx 1.56$, cf. Fig.~\ref{fig:psi_theta}b.
Parameters are $k_1=0.025$, $k_2=500$, $k_3=0.01$, $k_4=5$. 
The parameter $a_{stop}$ is taken as 20\% of the average amplitude of the 
autonomous system, i.e. before the feedback is turned on.

\subsection{An example: ensemble of chaotic R\"ossler oscillators}
With this example we demonstrate that the approach can be also applied to 
more complicated models and that suppression can be achieved with only two 
pulses per oscillatory cycle. Next, we explore the dependence of the 
performance on most important parameters.

We consider an ensemble of globally coupled chaotic R\"ossler oscillators:
\begin{equation}
\begin{cases}
\dot{x}_k &= -\w_k y_k -z_k +\e X + \cos\psi\cdot P(t)\;,\\
\dot{y}_k &= \w_k x_k +0.15y_k +\sin\psi\cdot P(t)\;,\\
\dot{z}_k &= 0.4+z_k(x_k-8.5)\;,
\end{cases}
\label{eq:ros}
\end{equation}
where frequencies $\w_k$ are Gaussian distributed with the mean $\w_0=1$ 
and standard deviation $0.02$. Without stimulation the system exhibits
the Kuramoto synchronization transition at the critical coupling 
$\e_{cr}\approx 0.05$ \cite{Pikovsky-Rosenblum-Kurths-96,Rosenblum-Pikovsky-04}. 
For $\e>\e_{cr}$ the mean-field dynamics is nearly periodic, while for $\e<\e_{cr}$ 
one observes small finite-size fluctuations of $X$.

\begin{figure}
\centering
\includegraphics[width=0.7\columnwidth]{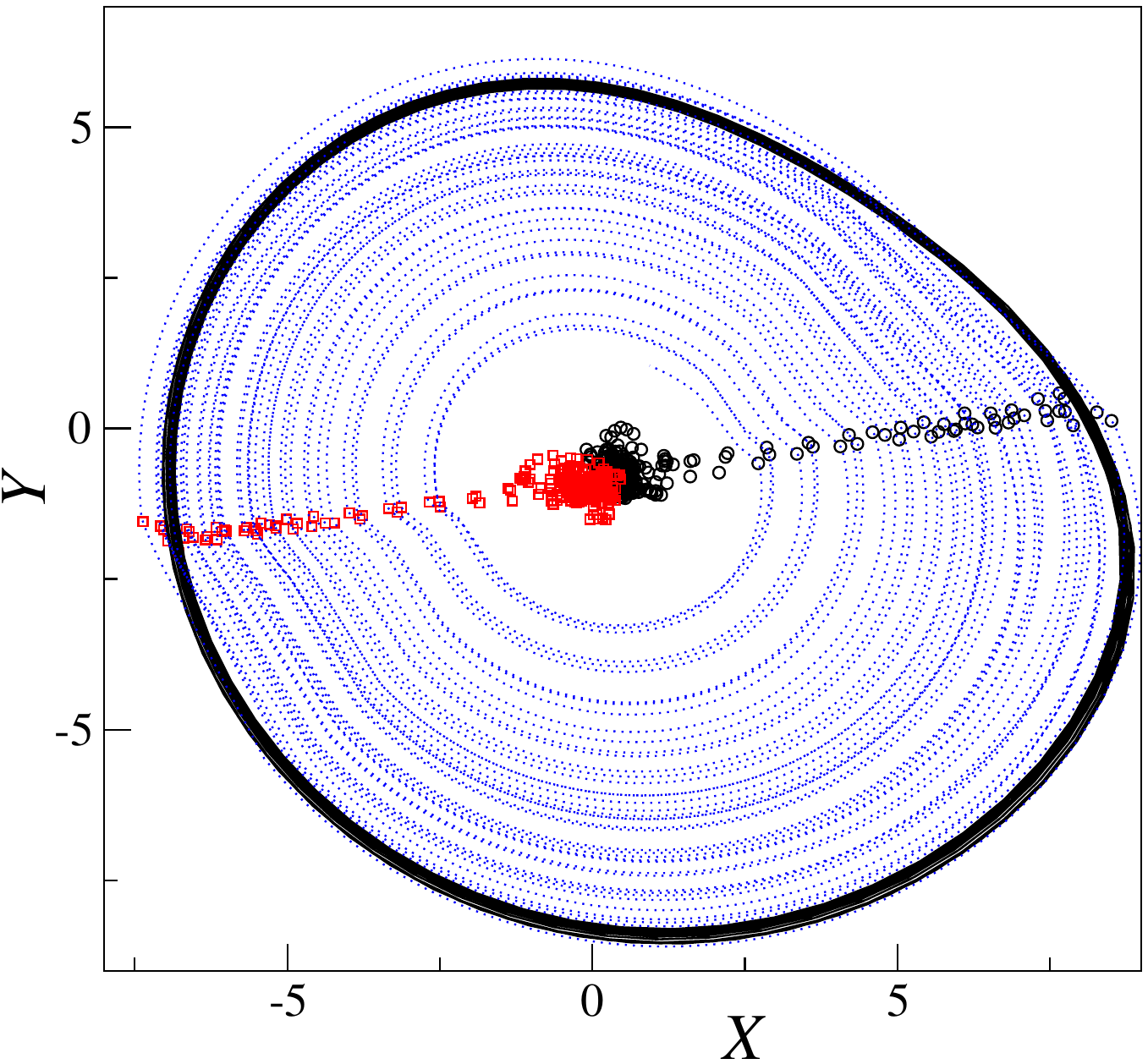}
\caption{Suppression of the collective mode in the ensemble of R\"ossler 
oscillators~(\ref{eq:ros}). Black solid curve shows a piece of trajectory
of the unforced system in the mean-field 
coordinates $X$ and $Y=N^{-1}\sum_k y_k$. 
Dashed curve shows trajectory of the controlled system.
(The trajectory is omitted for small amplitudes
for better visibility.)  
Black circles (red squares) indicate the points where negative (positive) pulses
are applied. 
}
\label{fig:ros1}
\end{figure}

First, in Fig.~\ref{fig:ros1} we illustrate suppression of synchrony in a system 
of $N=5000$ units, with $\e=0.1$ and $\psi=\pi/4$. 
Parameters of the feedback system are: $\w_0=1$, $k_1=0.001$, $k_3=0.001$,
$\Theta_{tol}=0.04\pi$, $A_0=2$, $\delta=0.2$, and $\Delta=0.4$ (other parameters are as 
given above). 
For the chosen value of $\Theta_{tol}$ only two pulses per cycles are applied (as 
can be seen in Fig.~\ref{fig:ros1}) and the adaptive algorithm converges to 
$\theta_0\approx 0.47$ and $\e_{fb}\approx -0.55$.~\footnote{Here, in order 
to shorten the transient we took $\e_{fb}(t_0)=-0.5$.} 
The suppression coefficient is $S=33.5$. 

\begin{figure}
\centering
\includegraphics[width=0.8\columnwidth]{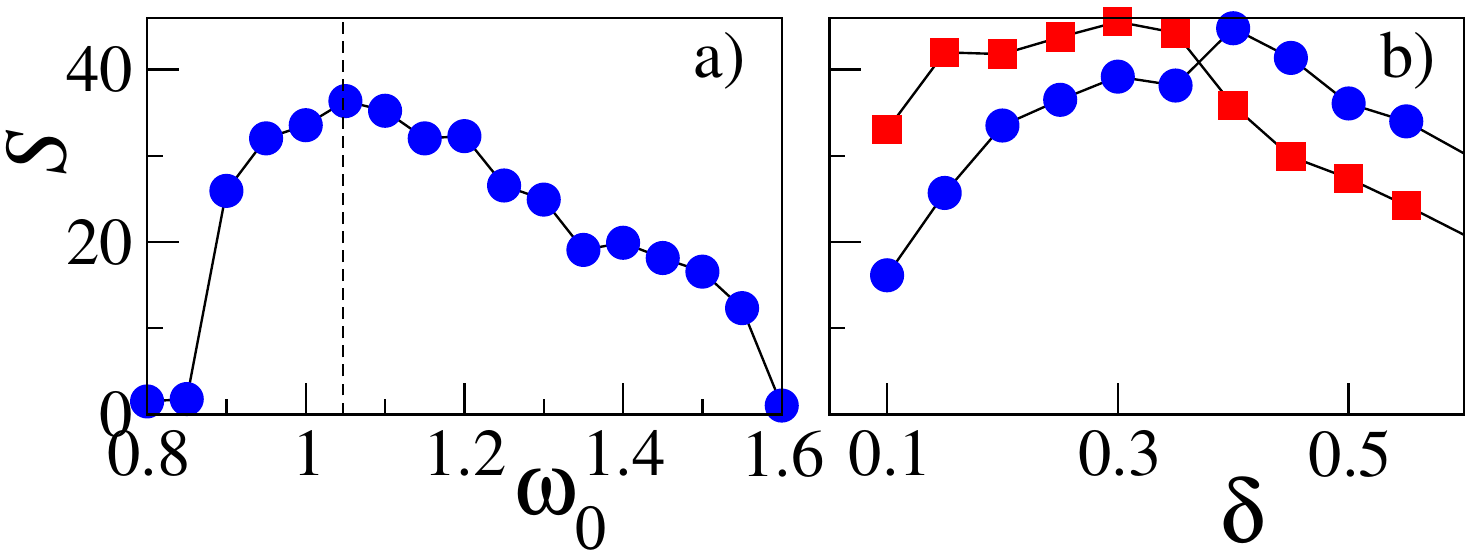}
\caption{Suppression coefficient $S$ in dependence on the frequency $\w_0$
of the linear oscillator Eq.~(\ref{eq:filter}) (a) and on the pulse width, $\delta$,
for $k_4=5$ (circles) and $k_4=0.5$ (squares). 
The vertical dashed line in panel (a) indicates the mean frequency of the unperturbed 
system's collective mode. 
The results show that the variation of $\w_0$ in the interval $\pm 15\%$ of the collective 
mode frequency provides a good suppression with $S\gtrsim 25$.}
\label{fig:ros2}
\end{figure}

Next, we check the dependence of $S$ on most important parameters, starting with the 
frequency of the linear oscillator, 
$\w_0$, see Eq.~\ref{eq:filter}.
Figure~\ref{fig:ros2}a presents the results. This plot demonstrates that the technique works for a rather broad 
range of $\w_0$. This feature is important for treatment of real-world systems with 
drifting average frequency.
The second test shows how the performance depends on the pulse width $\delta$ 
and on the parameter $k_4$ (Fig.~\ref{fig:ros2}b).
The latter determines saturation level for $\e_{fb}$, so that we can expect that 
the smaller $k_4$ the larger $\e_{fb}$ and, correspondingly, $S$. We also expect that 
broadening the pulse increases efficiency of suppression. Figure~\ref{fig:ros2}b
indicates that this expectation is correct unless the pulses become too wide and do not 
any more fit the interval of vulnerable phases.

Finally, we check whether the approach works for strongly 
coupled R\"ossler ensemble, $\e=0.2$ (see Eq.~\ref{eq:ros}).
For the pulse width $\delta=0.2$ and $k_4=5$ the technique fails, also 
with $A_0=4$. 
For $\delta=0.2$, $k_4=0.5$, and $A_0=4$ we achieve suppression with $S\approx 11$.
Helpful is also initial increase of the feedback, i.e. taking $\e_{fb}(t_0)=-1$, then the 
suppression works also with $k_4=5$. (We also compare the final values of $\e_{fb}$: 
for $k_4=5$ it remains $\approx 1$; for $k_4=0.5$ it tends to $-1.5$.)

Before proceeding, we briefly summarize how to choose other parameters of the algorithm.
The linear oscillator Eq.~(\ref{eq:filter}) acts as a bandpass filter and the 
damping factor $\alpha$ determines the width $\Delta f$ of the bandpass, 
$\Delta f=\alpha/2\pi$. Thus, if, e.g., the rhythm in question contains frequencies 
between 10 and 13 Hz, then $\Delta f$ shall be larger than 3 Hz, i.e., for this example 
$\alpha\gtrsim 0.3\omega_0$. 
Parameter of the integrating unit Eq.~(\ref{eq:integrator}) shall fulfill $\mu\gg 1$; 
the value $\mu=500$ ensures correct integration.

The stimulation parameter $\Theta_{tol}$ determines the number of stimuli in a bunch: 
the larger $\Theta_{tol}$, the larger is the number of stimuli, and, correspondingly, 
the suppression factor $S$, see Fig.~\ref{fig:results} below. On the other hand, the more stimuli 
in the bunch, the large is an intervention into the system. 
Hence, the optimal choice of $\Theta_{tol}$ depends on whether high values of 
$S$ or minimal intervention are preferred in a particular application.
The maximal stimulation amplitude $A_0$ in Eq.~(\ref{eq:pstrength}) also depends 
on the application: it shall be sufficiently small to ensure non-destructive action 
on the system. 

Finally, we discuss parameters $k_i$ in Eqs.~(\ref{adaptthe},\ref{adapteps}).
The product $k_1\bar a$ determines the adaptation step for the variable $\theta_0$. 
If this step is too big, then the adaptation rule may miss the optimal phase.  
If the step is too small, then the suppression is slow. 
A reasonable choice is to take $k_a\bar a\approx 0.01$. 
Similarly, the product $k_3\bar a$ determines the adaptation step for the feedback 
strength $\e_{fb}$, and this step also shall be neither too big nor too small. 
The value $k_3\bar a\sim  0.01$ works well with the tested models.
The choice of the parameter $k_2\gg 1$ is simple: it shall be big enough, 
e.g., $k_2=500$, so that the $\tanh$-function looks like the step function.

\section{Charged-balanced pulses}
\label{sec:chbal}
\begin{figure}
\centering
\includegraphics[width=0.8\columnwidth]{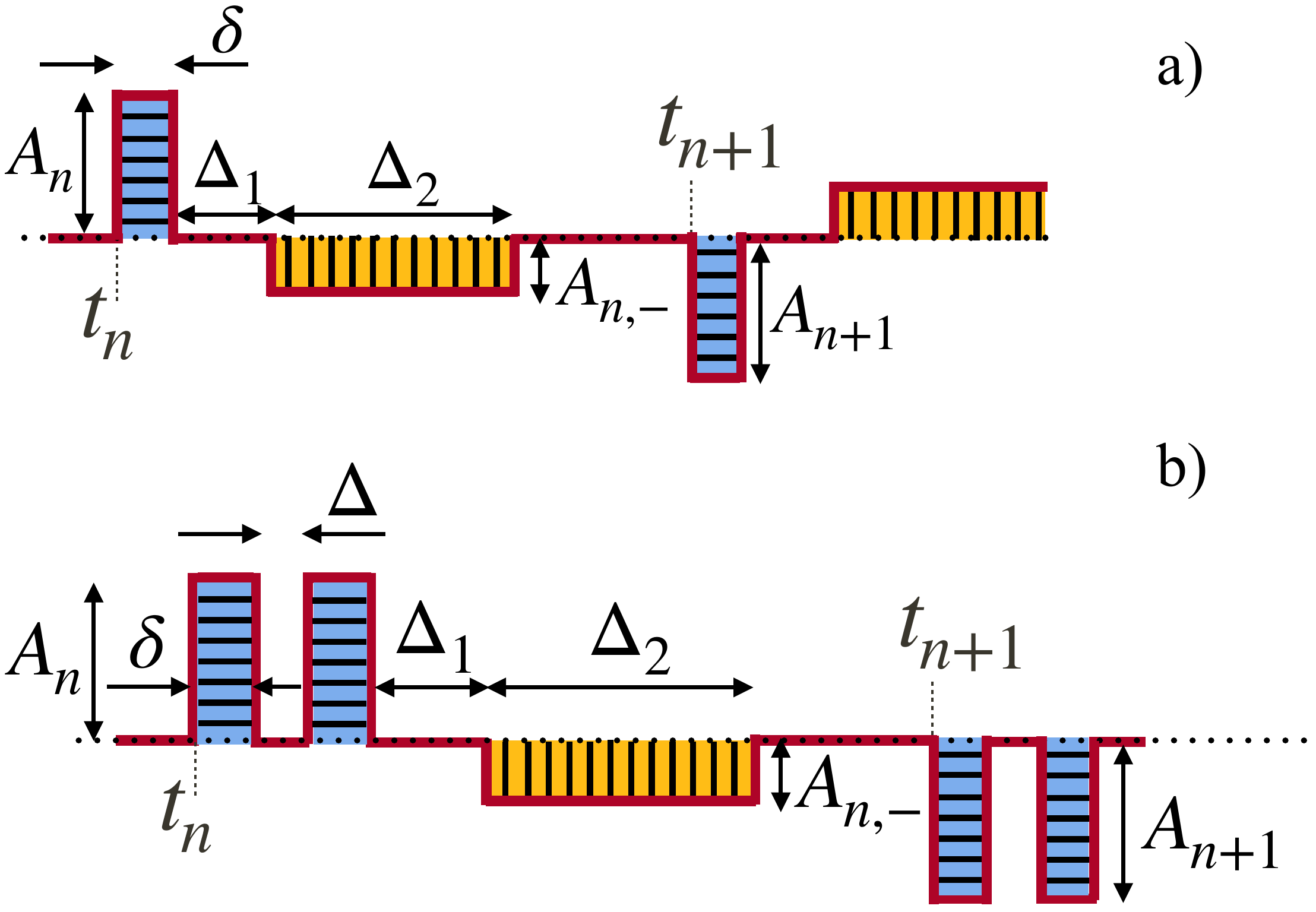}
\caption{Examples of bipolar charge-balanced stimuli. 
Panel (a) illustrates the simplest considered shape. 
Here two stimuli initiated at $t_n$ and $t_{n+1}$ are shown. Each stimulus 
consists of two rectangular pulses of opposite polarity and the 
charge-balance requirement means that the blue horizontally striped area equals 
the yellow vertically striped one, i.e. $A_n\delta=A_{n,-}\Delta_2$. 
Panel (b) illustrates a generalization of the shape in (a). 
Now $N_b$ narrow blue (horizontally striped) rectangular pulses are followed 
by one yellow (vertically striped) pulse of opposite polarity 
(the case $N_b=2$ is shown). The charge-balance condition becomes now
$N_bA_n\delta=A_{n,-}\Delta_2$.
}
\label{fig:pulse_shape2}
\end{figure}

The electrical stimulation of living systems requires a special form of pulses. 
Since the accumulation of electrical charge in the cells can be harmful, the pulses must be 
bipolar and charge-balanced. Figure~\ref{fig:pulse_shape2} provides the simplest example 
of such stimuli. In the rest of this Section, we explore desynchronization with 
charge-balanced stimulation.
For the test system, we again take the ensemble of globally coupled Bonhoeffer -- 
van der Pol oscillators, see Eqs.~(\ref{eq:bvdp}). 
If not said otherwise, the parameters are the same as in Section IIIb.

We begin with the stimuli shown in Fig.~\ref{fig:pulse_shape2}a. We fix $\delta=0.2$
and the amplitude ratio $A_n/A_{n,-}=-10$, then the charge-balance condition 
yields $\Delta_2=10\delta$. It turns out that the result of stimulation 
essentially depends on $\Delta_1$. If the negative part of the stimulus immediately 
follows the positive one (or vice versa) then their actions compensate each other.
Indeed, for $\Delta_1=0$ and $\Delta_1=2$ there is no suppression, $S\approx 1$
(see \cite{Popovych_et_al-17a} for a detailed model 
study on the suppression efficacy in dependence on the gap $\Delta_1$).
However, for $\Delta_1=6$, to be compared with the average oscillation 
period $T\approx 32.5$, the suppression factor is $S\approx 40$.  
It means that a narrow pulse comes in the vulnerable phase while the compensating 
wide pulse appears close to the least sensitive phase.
The efficiency of the suppression can improve if stimuli shown in 
Fig.~\ref{fig:pulse_shape2}b are used. We tested stimulation with 
$\delta=\Delta=0.2$, $\Delta_1=6$, and $N_b=2$ and $N_b=3$. As expected,
the maximal suppression with $S\approx 52$ is achieved for $N_b=3$.
Moreover, stimulation with $N_b=3$ and only once per period also succeeds 
to suppress the collective oscillation. 

\begin{figure}
\centering
\includegraphics[width=0.98\columnwidth]{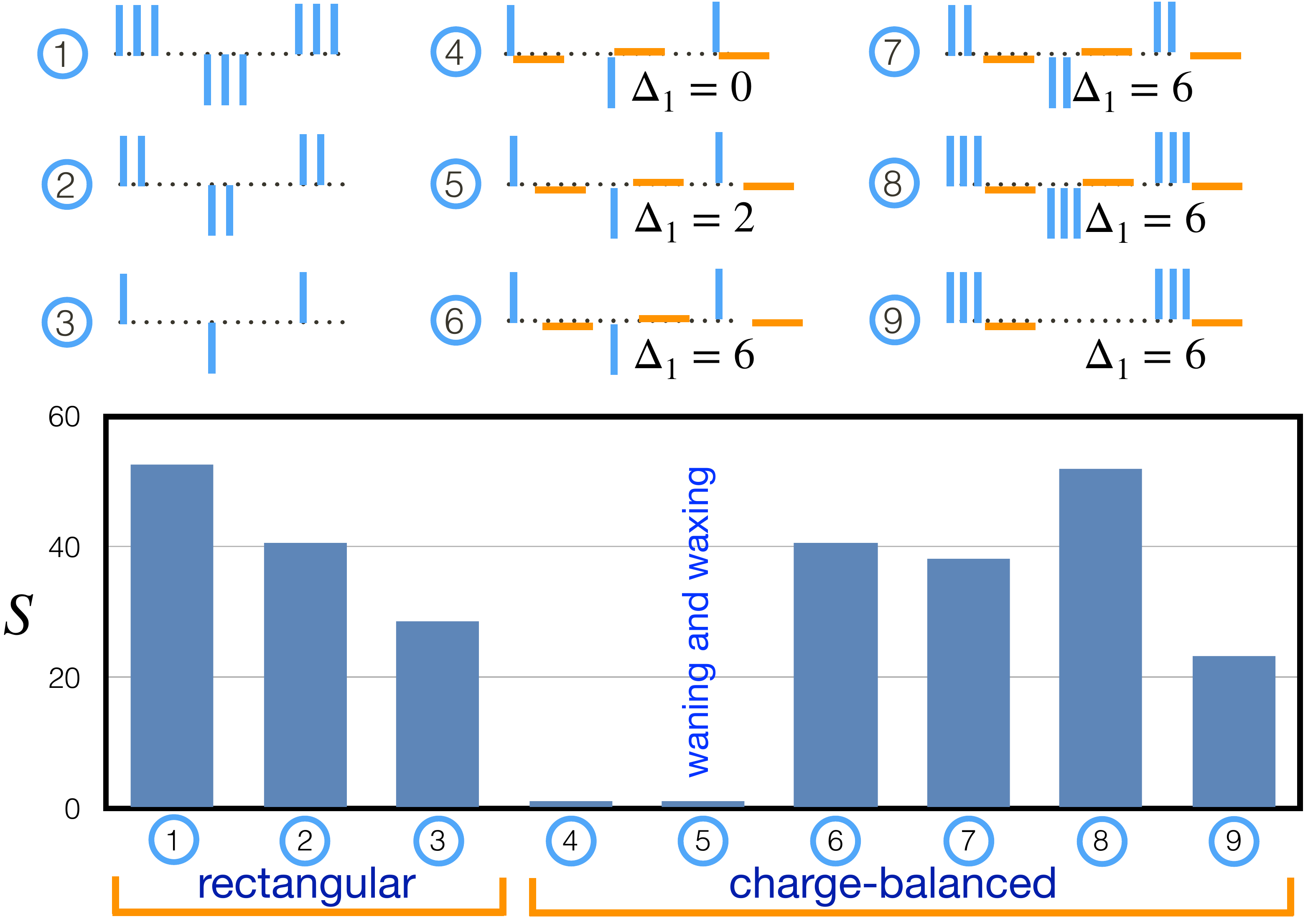}
\caption{Summary of the results for the Bonhoeffer -- van der Pol model. 
The suppression coefficient is shown for 9 different cases, presented schematically 
above the bar chart. 
Cases 1-3 and 4-9 correspond to stimulation by unipolar and charge-balanced pulses, respectively.
In cases 1-3 there are 3, 2, and 1 rectangular pulse in a burst around the vulnerable phase.  
Cases 4-6 correspond to stimulation with charge-balanced pulses with $N_b=1$ and 
a different delay between the narrow pulses and compensating wide one. 
In case 7 $N_b=2$, while in cases 8 and 9 $N_b=3$. 
Notice that in case 9 the stimulation is applied only once per period.
}
\label{fig:results}
\end{figure}

We summarize our results in the diagram shown in Fig.~\ref{fig:results}. 
For comparison, we present here both the results for simple rectangular as well as 
for charge-balanced pulses.  
In the trials 1-3 we exploit simple rectangular pulses. 
The tolerance parameter here is $\Theta_{tol}=0.08\pi$, $0.04\pi$, and $0.02\pi$,
what yields 3, 2, and 1 pulse in a burst, respectively. 
In cases 4-9 we use $\Theta_{tol}=0.02\pi$ and different values of $N_b$ and 
$\Delta_1$, see figure.
We notice that in case 6 one observes waning and waxing patterns. 
In fact, these patterns can be often obtained if the system is brought close to the border 
of suppression, by decreasing $\e_{fb}$.
This result might be interesting for neuroscience applications because some observations 
indicate that such regimes correspond to an improvement in the state of Parkinsonian 
patients~\cite{Tinkhauser17,*Tinkhauser18}.

\section{Enhancement of collective oscillation}
\label{sec:enh}
\begin{figure}
\centering
\includegraphics[width=0.5\columnwidth]{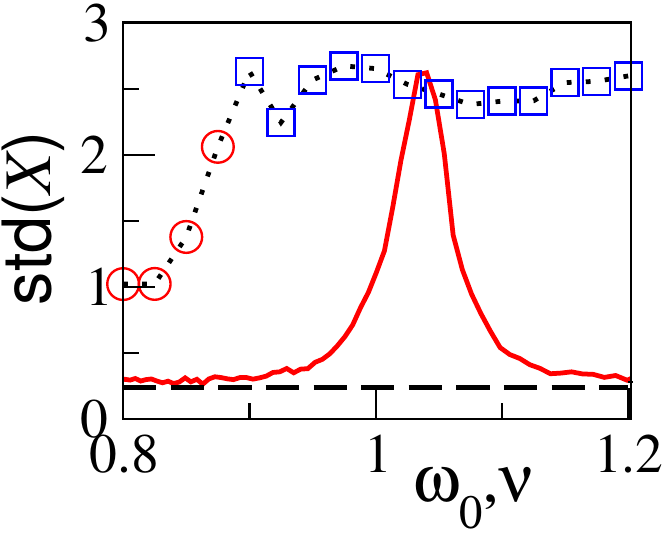}
\caption{Enhancement of the collective mode in the ensemble of R\"ossler
oscillators with the sub-threshold coupling. 
Red solid line shows dependence of $\sigma=\text{std}(X)$
on the frequency of the periodic pulsatile forcing $\nu$, while
the symbols illustrate the enhancement via the adaptive feedback-based 
approach with the rule Eq.~(\ref{adapthe2},\ref{adapteps2}); here $\sigma$ is plotted
vs. the parameter $\w_0$. Red circles and blue squares correspond to waning and waxing patterns 
and to stationary chaotic oscillation, respectively.
The dashed line indicates the level of finite-size fluctuations in the autonomous system.
}
\label{fig:enh}
\end{figure}

The simplest and most reliable way to increase ensemble synchrony by pulsatile 
stimulation is to apply the stimuli periodically, with some frequency $\nu$. 
This techniques is known as injection locking. 
It works for networks of periodic or chaotic oscillators, even if they are uncoupled.
However, the frequency $\nu$ shall be chosen in a proper way and this may be 
not an easy task if the frequency of the collective oscillation is not known beforehand 
and only finite-size fluctuations of the asynchronous ensemble are observed, 
especially in the presence of noise. 
The dependence of the standard deviation $\sigma$ of the mean field on the 
frequency of the drive has a typical resonance-like shape (see the solid line in 
Fig.~\ref{fig:enh}).
  
Enhancement can be also achieved via the feedback technique with slightly modified
update rules:   
\begin{align}
&\theta_0\;\to\; \theta_0+k_1 (A_{stop}-\bar A)(1+\tanh[k_2(A_{sat}-\bar A)]\;,\label{adapthe2}\\
&\e_{fb}\;\to\; \e_{fb}+k_3(A_{stop}-\bar A/)\cosh(k_4\e_{fb})\;,  \label{adapteps2}
\end{align}
where $A_{sat}$ is the saturation value. Certainly, this approach also has a frequency 
parameter, namely the frequency of the linear oscillator, $\w_0$. 
If the frequency of the collective oscillation is not known
\textit{a priori}, $\w_0$ shall be guessed. 
However, the results are not very sensitive
to the choice of $\w_0$, as illustrated in Fig.~\ref{fig:enh} 
for the model (\ref{eq:ros}) with the 
sub-threshold coupling $\e=0.02$. Other parameters are $A_0=1$, $A_{stop}=5$, 
$A_{sat}=2$. The rectangular pulses ($\delta=0.2$, $\Delta=1$) were used.
  
\bigskip
\section{Discussion and conclusions}
\label{sec:sum}
In summary, we presented and tested a closed-loop approach for control of collective activity in a 
globally coupled ensemble~\footnote{The approach certainly applies to a single oscillator as well. In this case, the stimulation quenches the self-sustained oscillation by keeping the system in a small vicinity of the unstable fixed point.}. 
The global coupling conjecture describes many natural phenomena, see, e.g., 
\cite{Strogatz-00,*Strogatz-03,Pikovsky-Rosenblum-Kurths-01,*Pikovsky-Rosenblum-15,%
Acebron-etal-05,*Osipov-Kurths-Zhou-07,*Breakspear-Heitmann-Daffertshofer-10}
and, in particular, models pathological brain activity in Parkinson's disease \cite{Tass-99,*Tass-00,*Tass-01,*Tass_2001,*Tass-02}. 
Global coupling is a good approximation for highly-interconnected networks, e.g., for randomly coupled neuronal networks \cite{Rosenblum-Tukhlina-Pikovsky-Cimponeriu-06}. 
Furthermore, numerical studies demonstrate that the feedback schemes developed for the globally coupled ensembles are efficient for controlling highly interconnected networks of excitatory and inhibitory neurons, e.g., of  STN and GPe cells with a realistic coupling 
scheme \cite{Popovych_et_al-17}.

The main advantage of our approach is that efficient control is accomplished by rare 
precisely timed pulses. 
So, we have shown that desynchronization can be achieved and maintained by only one stimulus per 
oscillatory cycle. The control parameters -- the feedback coefficient $\e_{fb}$ and the value of 
the phase $\theta_0$ when the system is most sensitive to stimulation -- are adjusted automatically. 
An essential feature of the approach is that the collective oscillation phase is estimated on the fly, 
using only previous values of the measured signal.  
The "device" for phase estimation is simple: it consists of a linear oscillator and integrator and,
therefore, can be easily implemented 
either via an electronic circuit or digitally. Another essential property of the feedback scheme 
is that it ensures the vanishing stimulation and maintains the desynchronized state by 
small-amplitude stimuli. 
Next, the developed technique aims at the desynchronization on the collective level while preserving the oscillation of individual units, i.e., the goal of the control is to avoid oscillation death of ensemble elements, but destroy the coherence of their activity. In this respect, our study follows the line of the research reported in \cite{Tass-99,*Tass-00,*Tass-01,*Tass_2001,*Tass-02,%
Rosenblum-Pikovsky-04,*Rosenblum-Pikovsky-04a,%
Popovych-Hauptmann-Tass-05,%
Tukhlina-Rosenblum-Pikovsky-Kurths-07,Hauptmann-Tass-09,*Popovych-Tass-12,Montaseri_et_al-13,%
Lin_2013,*Zhou_2017,*Wilson-Moehlis-16,*Holt_et_al-16,Popovych_et_al-17,Krylov-Dylov-Rosenblum-20}.  
All these studies assume 
that no access to individual units or connections is possible, neither for measurement nor for stimulation: it is supposed that only the collective mode can be observed and that control acts 
on the whole population, in contradistinction to the techniques like pinning or push-pull control 
\cite{Su-Wang-13,*He-Wang-Zhang-Zhan-14} that rely on access 
to a subset of the units.

The idea of applying pulses at a vulnerable phase goes back to P.A. Tass  publications \cite{Tass-99,*Tass-00,*Tass-01,*Tass_2001,*Tass-02} 
about twenty years ago.  The significant improvement brought by our approach is due to the feedback loop. Closed-loop control automatically detects the vulnerable phase and reduces the stimulus amplitude when the suppression is achieved.  As a result, the desired state is maintained by only one or two weak stimuli per oscillation period. Last but not least, our approach does not rely on phase approximation. 
Hence, it can be applied not 
only to ensembles of coupled limit-cycle oscillators but also to ensembles of chaotic units, e.g., 
bursting neuronal models. For the former case, when the phase sensitivity curve can be introduced, 
one can associate the vulnerable phase with the phase interval where the slope of the phase sensitivity 
curve is most steep. 
In this case, a common stimulation acts differently on units with close phases and shifts them apart, 
resulting in desynchronization.

Finally, we discuss the possible application of the proposed approach to neuroscience. 
We rely on a quite general assumption that the rhythms to be controlled emerge due to interaction in a 
large neuronal population. In this respect, we remain in the framework of the working hypothesis 
frequently exploited by the nonlinear community. 
Though we tested the approach on rather simple models, 
we believe that it works for more sophisticated ones, as long as the rhythms appear due 
to synchronization in a highly-interconnected network.  
Certainly, the dynamics of the human brain is much more complex, and the synchronization hypothesis 
may turn out too simplistic. However, our model-based approach can be useful as it is or in combination 
with ad hoc model-free closed-loop techniques for DBS that are nowadays under development in the 
neuroscience community~\cite{Rosin-11,*Little-13}. 
In this context we especially mention the phase-specific stimulation 
suggested and implemented 
in~\cite{Cagnan_at_al-13,*Cagnan_at_al-17,*Holt1119,%
*McNamara2020.05.21.102335,*Duchet_et_al-20}. 
We believe that incorporating the stimulus amplitude adaptation mechanism along with the automated tuning of phase relation shall be an essential improvement.

We emphasize several properties of our technique that render it suitable for DBS application. 
(i) It works with realistic charge-balanced stimuli. Though we have not searched for the optimal shape 
of stimuli, we have shown that stimulation is efficient if pulses of opposite polarity appear at the 
most and least sensitive phases, respectively, cf.~\cite{Popovych_et_al-17a}. 
It means that the condition of charge balance is fulfilled on a time scale of about one-fourth of 
the oscillatory cycle.  
(ii) Stimulation and measurement are separated in time. Indeed, the adaptive algorithm relies 
only on the values of the 
instantaneous amplitude between the epochs where stimulation is applied. 
(iii) Since the optimal phase for stimulation is determined automatically, it does not matter 
whether the mean field $X$  or its phase-shifted version is measured. 
This property is useful if the stimulation and measurement of brain activity are performed 
at different cites.
(iv) The linear oscillator used for phase estimation also acts as a bandpass filter and, therefore, 
extracts the rhythm of interest from the raw signal. However, the bandwidth of the filter is quite 
large -- the property required to deal with the signals with drifting frequency.

As a direction for further improvement, we mention the modification of the adaptation rule to allow 
for both increase and decrease of $\theta_0$. This modification will reduce the transient time for 
desynchronization and will help to avoid a temporal increase of synchrony in the process of adaptation.

\begin{acknowledgments}
The author acknowledges fruitful discussions with A.~Pikovsky, S.~Yanchuk, L.~Feldmann, A.~K\"uhn, 
and W.-J.~Neumann.
\end{acknowledgments}

\section*{Data availability}
Data sharing is not applicable to this article as no new data were created or analyzed in this study.

%

\end{document}